\RequirePackage{lineno}
\documentclass[a4paper]{jpconf}
\usepackage{graphicx}
\usepackage{verbatim}
\begin{document}
\title{Charged particle jet measurements with the ALICE experiment in
  proton-proton collisions at the LHC}
\author{S.~K.~Prasad for the ALICE Collaboration\footnote{for a full
    list of authors and the acknowledgements see~\cite{latestalice}.}}
\address{Wayne State University, 666 West Hancock Street, Detroit, MI, USA, 48201}
\ead{sprasad@cern.ch}
\begin{abstract}
  We present preliminary results of measurements of charged
  particle jet properties in proton-proton collisions at $\sqrt{s}$ =
  7 TeV using the ALICE detector. Jets are reconstructed using $\rm
  anti-k_{T},~k_{T}$ and SISCone jet finding algorithms with resolution
  parameter $R=0.4$ in the range of 
  transverse momentum from 20 to 100 GeV/$c$ in the
  midrapidity region ($\mid\eta\mid\textless$
  0.5). The uncorrected charged jet spectra obtained using the three
  different jet finders show good agreement. The data are compared to
  predictions from
  PYTHIA-Perugia0, PYTHIA-Perugia2011, and PHOJET.
  The mean charged particle multiplicity in leading jets increases with 
  increasing jet $p_{\rm T}$ and is consistent with model
  predictions. The radial distributions of transverse
  momentum about the jet
  direction and  the distributions of the average radius containing 80\%
  of the total jet $p_{\rm T}$ found in the jet cone ($R = 0.4$ in this analysis),
  indicate that high 
  $p_{\rm T}$ jets are more collimated than low $p_{\rm T}$ jets.  
\end{abstract}
\section{Introduction}
Jets are collimated sprays of particles originating from the fragmentation of
hard scattered partons (quarks and gluons) produced in pp (or A--A)
collisions~\cite{JetDef}. As such, jets serve as a proxy for the high
$p_{\rm T}$ partons produced in elementary 
hard scatterings. In pp collisions, they provide a unique tool to test 
perturbative quantum chromodynamics (pQCD)~\cite{RefpQCD}.
Measurements of  jet shapes in particular, provide the details of the
jet fragmentation process~\cite{JetDef}. Jet shapes are also sensitive
to the type of partons (quarks or gluons) that fragment into hadrons.
In addition, measurements in pp collisions
provide a baseline for similar measurements in 
more complex A--A collisions where various nuclear effects are
expected to take place. Jet shape observables have previously been
measured by the CDF~\cite{cdfprl70, cdfprd65, cdfprd71} and
D0~\cite{d0plb357} collaborations in $\rm p\bar{p}$ collisions and
more recently by the ATLAS~\cite{atlasprd83} and CMS~\cite{cmsanalysis}
collaborations in pp collisions at $\sqrt{s}$ = 7 TeV.
In this paper we report preliminary results of charged particle
jet properties measured by A Large Ion Collider Experiment (ALICE) in
pp collisions at $\sqrt{s}$ = 7 TeV. 
\section{Data analysis}
The data used in this analysis were collected during the
Large Hadron Collider (LHC)~\cite{lhccdr_1,lhccdr_2} run in the year 2010 with the ALICE
detector~\cite{AliceExpt}. 
The main detector subsystems used for tracking are the Time Projection
Chamber (TPC)~\cite{RefTPC} and the Inner Tracking System
(ITS)~\cite{RefITS}. 
The V-ZERO (V0) counters~\cite{RefVzero}
and the ITS are used for online trigger to select the minimum bias
events. 
Only events
with primary vertex within $\pm$10 cm along the beam axis from the nominal interaction point are
analyzed to ensure uniform acceptance in pseudo-rapidity
($\eta$) and minimize the need for complex acceptance corrections.
This analysis is based on 161 M minimum bias
events. Tracks are reconstructed using combined information from the
TPC and the ITS. These tracks have uniform high tracking efficiency for charged
particles, good momentum resolution and uniform $\phi$ distribution. Tracks are
included in the analysis if $p_{\rm T, track}\textgreater$ 0.150 GeV/$c$
and $\mid\eta_{\rm track}\mid\textless$ 0.9.
\section{Jet reconstruction method and observable definitions}
Charged particle jets are reconstructed using the sequential
recombination algorithms anti-$\rm k_{T}$~\cite{RefAntikt} and $\rm
k_{T}$~\cite{RefKt_1, RefKt_2, RefKt_3} from the FastJet package~\cite{RefFastjet} and a
seedless infrared-safe cone based SISCone~\cite{RefSiscone}
algorithm. Jets are reconstructed with the resolution parameter $R =
0.4$. Only jets within $\mid\eta\mid$ $\textless$ 0.5 and in the
transverse momentum range from 20 to 100 GeV/$c$ are considered in this
analysis. The jet properties are measured for leading jets
only i.e. the jet with the highest $p_{\rm T}$ in each event. The
jet shape observables studied in this analysis are defined in the
following subsections.
\subsection{Charged particle multiplicity in leading jet}
The charged particle multiplicity of the leading jet is defined as the
number of charged particles within the jet
cone. The mean is computed in bins of the jet $p_{\rm T}$. 
\subsection{Leading charged jet size $(R_{\rm 80})$}
The size of the jet, $R_{\rm 80}$, is defined as the radius in $\rm
\eta-\phi$ space that contains 80\% of the total transverse momentum
found in the jet cone ($R = 0.4$ in this analysis). We report the mean
value of $R_{\rm 80}$ for
leading jets as a function of jet $p_{\rm T}$.
\subsection{Radial distribution of transverse momentum within the
  leading charged jet}
\begin{figure}[here]
\centering
\includegraphics[width=5.5cm,height=5.5cm]{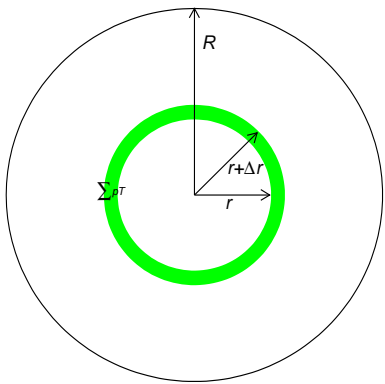}
\caption{\small{Illustration of radial distribution of scalar
   $p_{\rm T}^{\rm sum}$ about
   the jet axis in the leading charged jet. Scalar $p_{\rm T}^{\rm
     sum}$ is obtained for 
   each annular region as a function of distance $r =
   \sqrt{(\Delta\eta)^{\rm 2} + (\Delta\phi)^{\rm 2}}$ from the jet axis.}}
\label{cartoon}
\end{figure}

We also study the distribution of transverse momentum within a leading
charged jet as a function of distance
$r = \sqrt{(\Delta\eta)^{\rm 2}+(\Delta\phi)^{\rm 2}}$ from the jet
direction. 
We construct annular regions as illustrated in
Fig.~\ref{cartoon} at given $r$ values. The scalar sum of the transverse
momenta ($p_{\rm T}^{\rm sum}$) of all charged particles produced in each
annulus is calculated jet by jet as a function of distance $r$. 
We report the mean value of this sum (Eq.~\ref{eq1}) as a function
of jet $p_{\rm T}$.
\begin{equation}\label{eq1}
 <\frac{dp_{\rm T}^{\rm sum}}{dr}>(r) = \frac{1}{\Delta
   r}\frac{1}{\rm N_{jets}}\sum p_{\rm T}(r-\Delta r/{\rm 2},
  r+\Delta r/{\rm 2})
\end{equation}
where, $p_{\rm T} (r-\Delta r/{\rm 2}, r+\Delta r/{\rm 2})$ denotes the summed
$p_{\rm T}$ of all tracks inside the annular ring between $r-\Delta
r/{\rm 2}$
and $r+\Delta r/{\rm 2}$. $\Delta r$ is the radial width of the
annulus. The analysis was carried out with $\Delta r$ = 0.02. $\rm
N_{jets}$ denotes the 
total number of jets.
\section{Instrumental effects and systematic uncertainties}
We used a bin-by-bin correction procedure similar to those used by
the CDF~\cite{cdfprd65} and ATLAS~\cite{atlasprd83} collaborations to
correct the measured jet shape observables for detector effects. 
PYTHIA 6.4~\cite{pythia6p4} Monte Carlo (MC) with tune
Perugia0~\cite{perugia0} 
describes the measured jet shapes reasonably well for
$R = 0.4$ and therefore is used for the correction. Correction factors are
computed for each bin in jet $p_{\rm T}$. They are 
defined as the ratio between the jet shape observable at particle
level to that at detector level in their
respective jet $p_{\rm T}$ bin as described by Eq.~\ref{eq2},
\begin{equation}\label{eq2}
  \rm  CF(p_{t}) = \frac{Obs_{mc}^{part}}{Obs_{mc}^{det}}
\end{equation}
where, $\rm Obs_{mc}^{part}$ is the observable
at particle level and $\rm Obs_{mc}^{det}$ is that at the detector level.
Jet shape observables at particle level are
obtained using MC jets at particle level. A full
GEANT~\cite{RefGeant3} simulation
implementing the detector effects has been performed
to obtain the observables at detector level. We have not corrected for
underlying events (UE).

The main contributors to the systematic uncertainties are the
uncertainties in efficiency and momentum resolution in the track
reconstruction. These effects were estimated using MC (PYTHIA) events.
Uncertainties on the measured observables are estimated
by varying the efficiency and momentum resolution by 5\% and 20\% respectively. 
\section{Results}
\subsection{Comparative performance analysis of jet finders}
\begin{figure}[here]
\centering
\includegraphics[height=6.5cm, width=8.5cm]{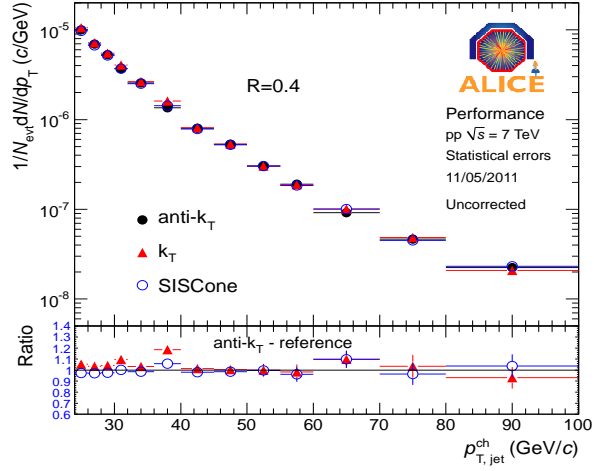}
\caption{(Top panel) Uncorrected $p_{\rm T}$-distributions of jets
  for pp collisions at $\sqrt{s}$ =
  7 TeV obtained with the $\rm anti-k_{T}$ (filled black circles),
  $\rm k_{T}$ (filled red
  triangles), and SISCone (open blue circles) jet
 finding algorithms. (Bottom panel) The ratios between the yields
 obtained with the $\rm k_{T}$, and SISCone to that with the
 $\rm anti-k_{T}$. The  solid line at unity is to guide the
 eye. Error bars indicate the statistical uncertainties.}
\label{dndpt4method}
\end{figure}

Figure~\ref{dndpt4method} (Top panel) shows the uncorrected transverse
momentum distributions of charged particle jets obtained using the
$\rm anti-k_{T}$
(filled black circles), $\rm k_{T}$ (filled red triangles),
and SISCone (open blue circles) jet finding algorithms with $R = 0.4$
within $\mid\eta\mid\textless$ 0.5, in the range of
transverse momentum from 20 to 100 GeV/$c$ for pp collisions at
$\sqrt{s}$ = 7 TeV. 
The ratios between the uncorrected yields
obtained with the $\rm k_{T}$, and SISCone to that
obtained with the anti-$\rm k_{T}$ are shown in the bottom
panel. We observe that the three different jet reconstruction
algorithms agree well with each other at detector level.
\subsection{Charged particle multiplicity in leading jet}
The mean charged particle multiplicity ($\rm \textless~N_{ch}~\textgreater$)
\begin{figure}[here]
  \centering
  \includegraphics[height=6.5cm, width=8.5cm]{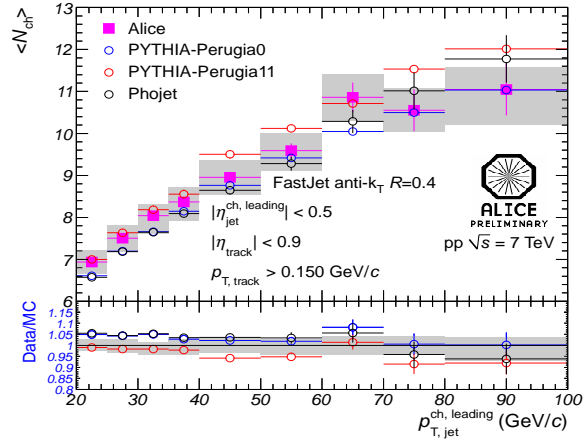}
  \caption{(Top panel) Mean charged particle multiplicity in the leading jet as a
    function of jet $p_{\rm T}$ for pp collisions at $\sqrt{s}$ = 7 TeV
  (filled magenta squares) compared to predictions from PYTHIA-Perugia0~\cite{perugia0}
  (open blue circles), PYTHIA-Perugia2011~\cite{perugia0} (open red circles) and
  PHOJET~\cite{phojet} (open black circles). (Bottom panel) The ratios between
  data to MC predictions. Error bars indicate the statistical uncertainties while the
  gray bands indicate the systematic uncertainties.}
\label{nCh}
\end{figure} 
distributions for leading jets and their dependence on jet $p_{\rm T}$ are shown in Fig.\ref{nCh}.
One observes that the mean charged multiplicity increases with
increasing jet $p_{\rm T}$
as expected based on prior measurements by the CDF~\cite{cdfprd65}
collaboration. This behavior is observed also in the MC
predictions. The ratios of measured data to MC predictions suggest
that models reproduce the data rather well.
\subsection{Leading charged jet size}
\begin{figure}[here]
  \centering
  \includegraphics[height=6.5cm, width=8.5cm]{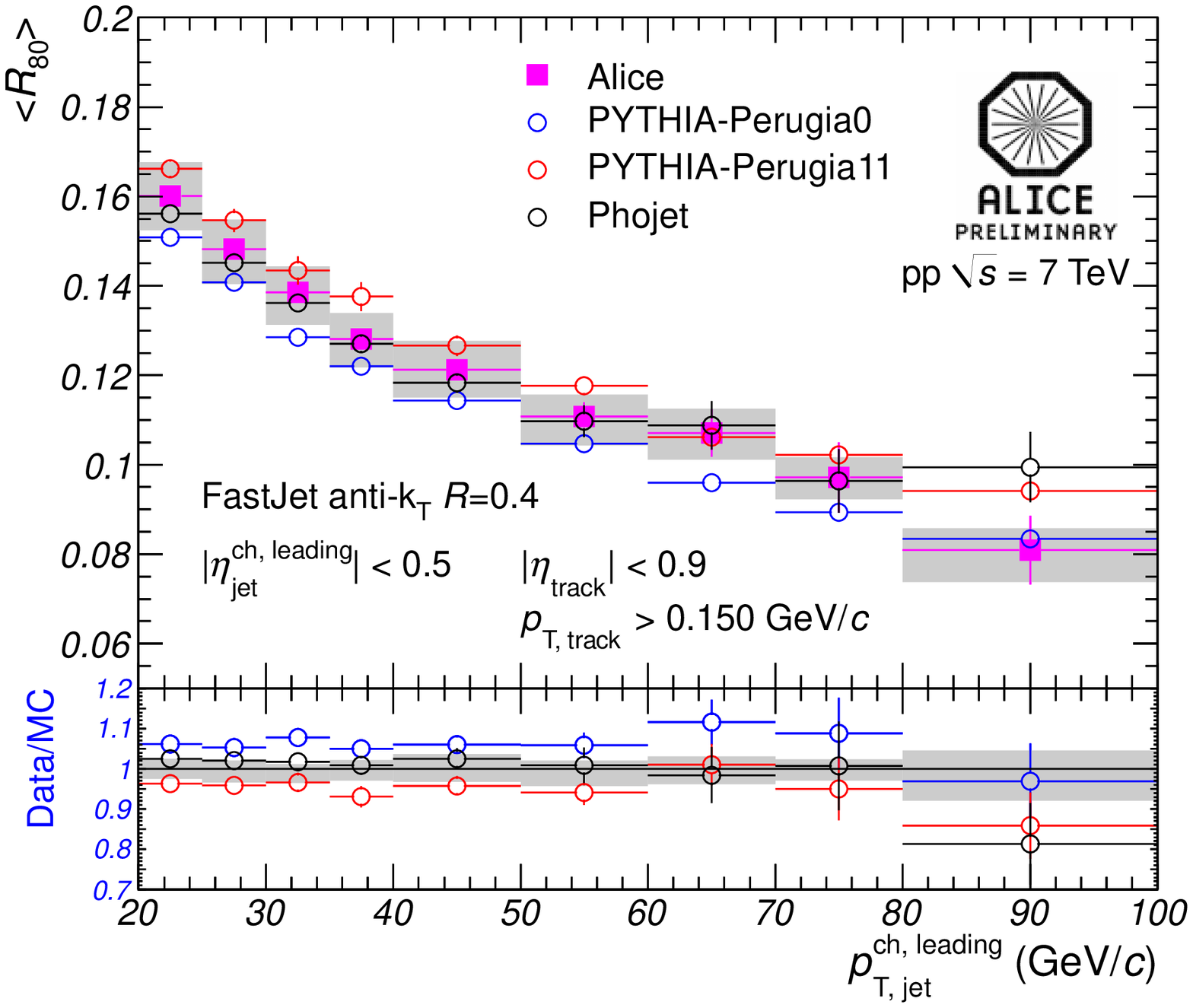}
  \caption{(Top panel) Distributions of the average radius $R_{\rm 80}$ containing
    80\% of jet $p_{\rm T}$ found in the jet cone ($R = 0.4$) as a function of
    jet $p_{\rm T}$ for pp
    collisions at $\sqrt{s}$ = 7 TeV (magenta squares) compared to predictions
    from PYTHIA-Perugia0~\cite{perugia0} (open blue circles),
    PYTHIA-Perugia2011~\cite{perugia0} (open
    red circles) and PHOJET~\cite{phojet} (open black circles). (Bottom panel) The
    ratios between the data to the predictions from simulations. Error bars indicate the statistical uncertainties while the
    gray bands indicate the systematic uncertainties.}
  \label{r80}
\end{figure}   
The distributions of the average radius ($\textless~R_{\rm
  80}~\textgreater$) containing 80\% of the total jet
$p_{\rm T}$ found in the jet cone ($R = 0.4$) are shown in Fig.~\ref{r80} as a
function of jet $p_{\rm T}$.
At low jet $p_{\rm T}$ (20 GeV/$c$), $\rm <R_{\rm 80}>$ is contained in a
very small cone of radius $\sim$0.16, 
which decreases further down to $\sim$0.08 for high $p_{\rm T}$ jets
(100 GeV/c). 
\subsection{Radial distribution of transverse momentum within the
  leading charged jet}
\begin{figure}[h]
  \begin{minipage}{19pc}
    \includegraphics[height=6.5cm, width=19pc]{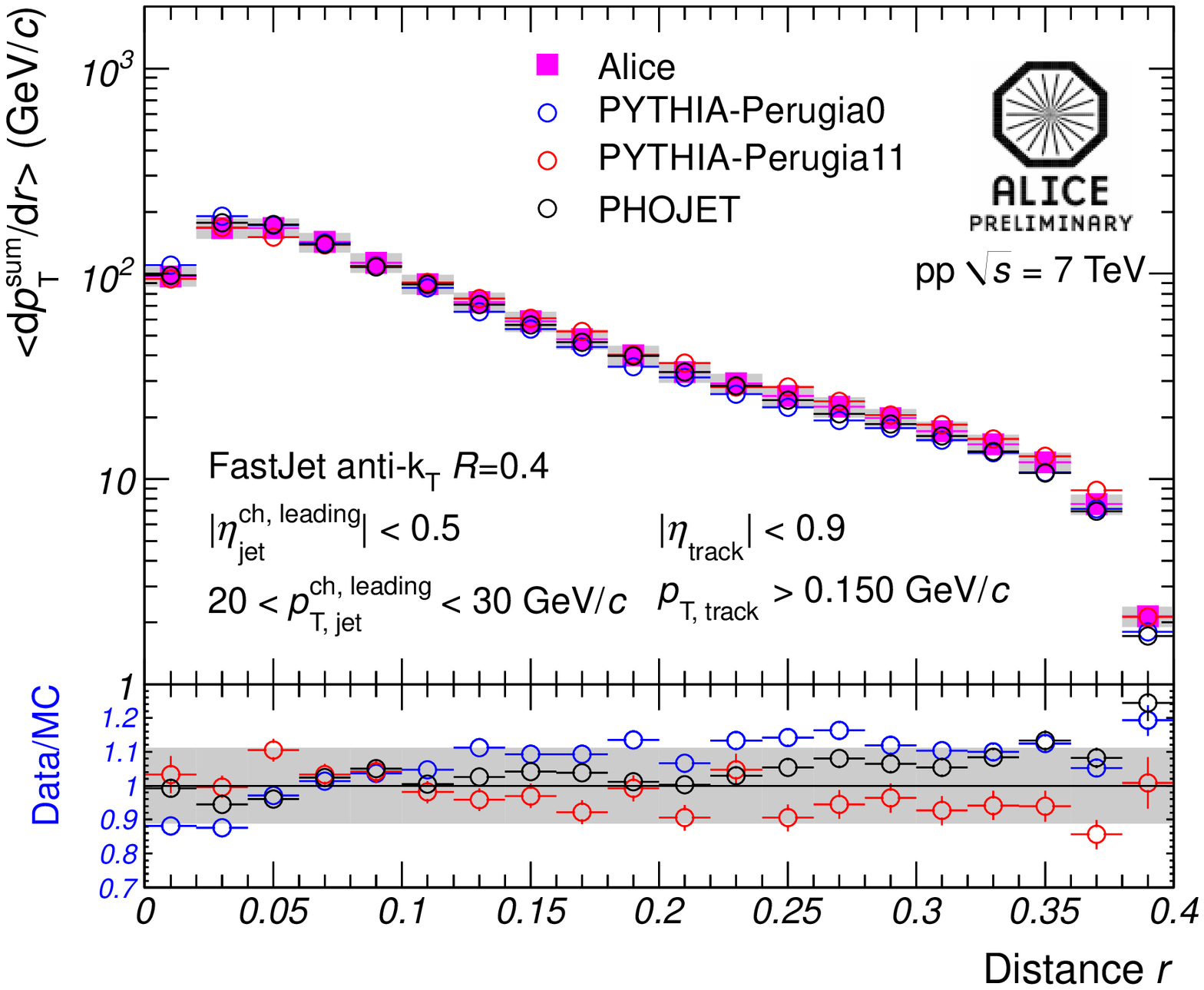}
  \end{minipage}
  \begin{minipage}{19pc}
    \includegraphics[height=6.5cm, width=19pc]{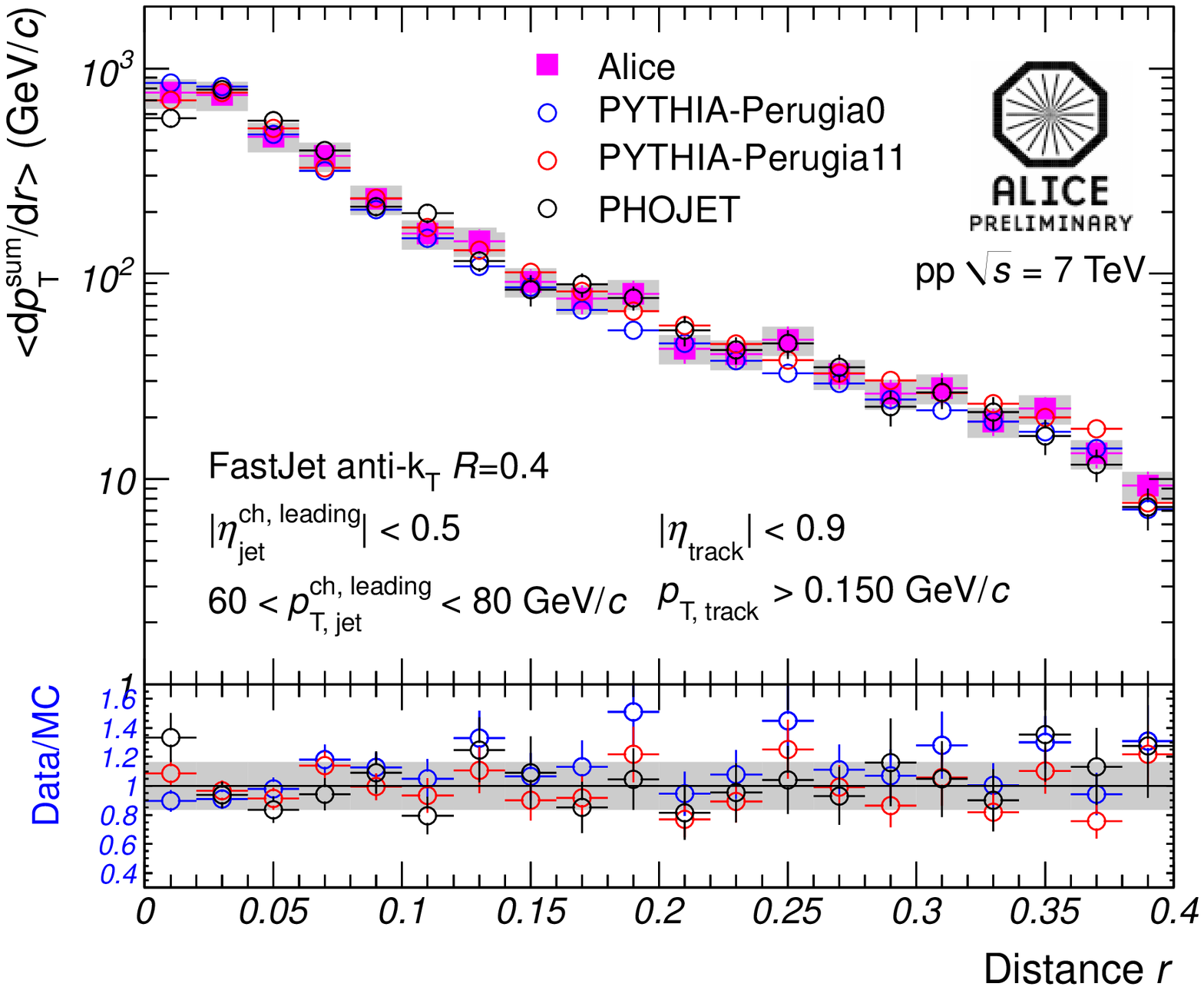}
  \end{minipage}
  \caption{(Left panel) (Top) Radial distributions of transverse momentum about
    the jet axis in the leading charged jet (magenta squares) for pp
    collisions at $\sqrt{s}$ = 7 TeV for 20
    $\textless~{\rm jet}~p_{\rm T}~\textless$ 30 GeV/$c$ compared to predictions
    from PYTHIA-Perugia0~\cite{perugia0} (open blue circles),
    PYTHIA-Perugia2011~\cite{perugia0} (open
    red circles) and PHOJET~\cite{phojet} (open black circles). (Bottom) The
    ratios between the data to the predictions from simulations. (Right panel)
    Results for 60 $\textless~{\rm jet}~p_{\rm T}~\textless$ 80
    GeV/$c$. Error bars indicate the statistical uncertainties while the
    gray bands indicate the systematic uncertainties.}
  \label{raddist}
\end{figure}
Figure~\ref{raddist} shows the radial distributions of transverse
momentum in the leading jet about the jet axis for jet $p_{\rm T}$ between
20 to 30 GeV/$c$ (left panel) and 60 to 80 GeV/$c$ (right panel).
It can be clearly seen in the figure that the transverse momentum
density is largest near the jet axis and decreases with $r$. 
The higher slop for
this decrease in case of high $p_{\rm T}$ jets
(right panel) indicates that high $p_{\rm T}$ jets are more
collimated than low $p_{\rm T}$ jets. 
\section{Conclusions}
We reported preliminary measurements of charged particle jet
properties in pp collisions at $\sqrt{s}$ = 7 TeV using the
ALICE detector. Jets were reconstructed using the $\rm anti-k_{T}$,
$\rm k_{T}$, and
SISCone jet finding algorithms with
resolution parameter $R = 0.4$ in the $p_{\rm T}$ range from 20 to 100
GeV/$c$. A comparison between the uncorrected charged jet spectra
obtained using three different jet finding algorithms show a good
agreement. The mean charged
particle multiplicity in the leading charged jet increases with increasing
jet $p_{\rm T}$. Similar to the previous observations reported by
the CDF~\cite{cdfprd65} and ATLAS~\cite{atlasprd83} collaborations,
results on $R_{\rm 80}$ and $p_{\rm T}^{\rm sum}$ show
that high $p_{\rm T}$ jets are more collimated than low $p_{\rm T}$
jets. Predictions from simulations (PYTHIA-Perugia0,
PYTHIA-Perugia2011, and PHOJET) describe the jet shape observables
reasonably well.
\section*{References}

\end{document}